\documentclass[conference,hidelinks]{IEEEtran}
\IEEEoverridecommandlockouts

\usepackage{cite}
\usepackage{amsmath,amssymb,amsfonts}
\usepackage{algorithmic}
\usepackage{graphicx}
\usepackage{textcomp}
\usepackage{xcolor}
\def\BibTeX{{\rm B\kern-.05em{\sc i\kern-.025em b}\kern-.08em
    T\kern-.1667em\lower.7ex\hbox{E}\kern-.125emX}}

\usepackage{hyperref}

\newcommand{\orcid}[1]{\href{https://orcid.org/#1}{\includegraphics[height=10pt]{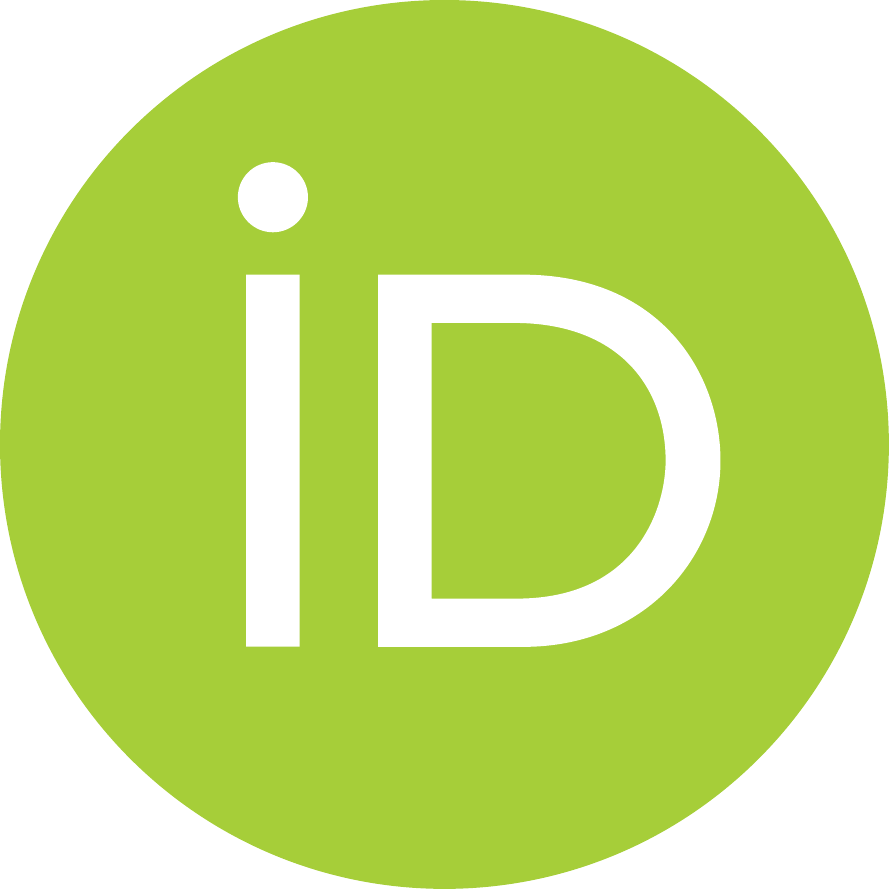}}}

\usepackage{tikz}
\usepackage{pgfplots}

\usepackage{siunitx}
\usepackage[caption=false, font=footnotesize]{subfig}

\usepackage{aas_macros}

\usepackage{xcolor}
\definecolor{plot1}{HTML}{003f5c}
\definecolor{plot2}{HTML}{bc5090}
\definecolor{plot3}{HTML}{ffa600}

\usepackage{booktabs}
\usepackage{todonotes}

\usepackage{listings}

\begin{document}
\bstctlcite{IEEEexample:BSTcontrol}

\title{Distributed, combined CPU and GPU profiling within HPX using APEX
}

\author{
\IEEEauthorblockN{Patrick Diehl\orcid{0000-0003-3922-8419}\IEEEauthorrefmark{1}\IEEEauthorrefmark{4}, Gregor Dai\ss\IEEEauthorrefmark{2}, Kevin Huck\orcid{0000-0001-7064-8417}\IEEEauthorrefmark{3}, Dominic Marcello\IEEEauthorrefmark{1} , Sagiv Shiber\orcid{0000-0001-6107-0887}\IEEEauthorrefmark{4}, Hartmut Kaiser\orcid{0000-0002-8712-2806}\IEEEauthorrefmark{1},\\ Juhan Frank\IEEEauthorrefmark{4}, Geoffrey C. Clayton\orcid{0000-0002-0141-7436}\IEEEauthorrefmark{4}, and  Dirk Pfl\"uger\orcid{0000-0002-4360-0212}\IEEEauthorrefmark{2}}
\IEEEauthorblockA{
\IEEEauthorrefmark{1}LSU Center for Computation \& Technology, Louisiana State University,
Baton Rouge, LA, 70803 U.S.A\\
Email: patrickdiehl@lsu.edu}
\IEEEauthorblockA{\IEEEauthorrefmark{2} IPVS, University of Stuttgart,
Stuttgart, 70174 Stuttgart, Germany}
\IEEEauthorblockA{\IEEEauthorrefmark{3} \textit{OACISS}, University of Oregon, Eugene, OR, U.S.A.}
\IEEEauthorblockA{\IEEEauthorrefmark{4} Department of Physics and Astronomy, Louisiana State University,
Baton Rouge, LA, 70803 U.S.A. }
}


\maketitle

\begin{abstract}
Benchmarking and comparing performance of a scientific simulation across hardware platforms is a
complex task.  When the simulation in question is constructed with an asynchronous, many-task  (AMT)
runtime offloading work to GPUs, the task becomes even more complex.
In this paper, we discuss the use of a uniquely suited performance measurement library, \textit{APEX},
to capture the performance behavior of a simulation built on \textit{HPX}, a highly scalable, distributed
AMT runtime.  We examine the performance of the astrophysics simulation carried-out by \textit{Octo-Tiger}
on two different supercomputing architectures.  We analyze the results of scaling and measurement
overheads. In addition, we look in-depth at two similarly configured executions on the two
systems to study how architectural differences affect performance and identify opportunities
for optimization. 
As one such opportunity, we  optimize the communication for the hydro solver and investigated its performance impact.
\end{abstract}


\section{Introduction}
\label{sec:intro}
Whether in CPU code, in GPU kernels, or in the inter-node communication --
performance bottlenecks in High-Performance Computing (HPC) applications may be
hidden in any part of the program.
There have been many attempts to ease the development of HPC applications and
to indicate and to avoid as many bottlenecks as possible.
One example is the asynchronous many-task (AMT) system HPX~\cite{Kaiser2020},
which aims to solve some of the more common problems. It makes it easier to
overlap communication and computation, to employ both CPUs and GPUs, and to
avoid overhead for fine-grained parallelism using lightweight threading. 

Even with HPX, of course, it is still perfectly possible to introduce
performance bottlenecks into one's application. Being able to collect
performance measurements to profile an HPX application remains important. For
AMT systems such as HPX, it is beneficial to have a profiling tool that
understands the task-based nature of the runtime system -- for example, call stacks
themselves are less useful: A system thread may jump back and
forth between various HPX tasks as they are yielded and resumed, and the call
stack itself may be dozens of levels of runtime functions that have no
particular interest to the application developer.

Collecting these measurements in a profiling run is challenging, as one not
only needs a profiling framework that supports both CPU and GPU code as well as
the distributed collection of profiling data across many compute nodes. One
also needs to keep any overheads introduced by the profiling itself to an
absolute minimum. Otherwise, it would not only distort the collected
measurements, but also make large, distributed runs infeasible, rendering us
unable to detect potential performance bottlenecks that only appear at scale. 

HPX is integrated with a performance measurement library, APEX (Automatic
Performance for Exascale), which was designed specifically for the HPX
runtime and the above requirements.
In previous work, APEX was used together with the HPX performance counters to
collect performance measurements for Octo-Tiger \cite{marcello2021octo}, an
astrophysics application which is built upon HPX and contains optimized kernels
for both CPUs and GPUs~\cite{pfander2018accel,daiss2021beyond}. Octo-Tiger is
capable of running the same kernels on the CPUs and GPUs simultaneously (on
different data) depending on the load. In that work, profiling data of
Octo-Tiger was gathered with APEX in distributed CPU-only runs where the energy
usage, the idle rate, and overhead of the HPX AGAS (Active Global Address
Space) was analyzed~\cite{diehl2021performance}.
Furthermore, combined CPU-GPU profiling runs have been performed on Summit,
analyzing the performance behavior of Octo-Tiger's new CUDA\textsuperscript{\texttrademark} hydro module in
different configurations for simple benchmark
scenarios~\cite{diehl2021octotigers}.


All these previous efforts inspire this new work, collecting performance
measurements on both CPU and GPU during a full-scale production-scenario run on
Piz Daint (Cray\textsuperscript{\texttrademark} XC50 with one 12-core Intel\textsuperscript{\textregistered} Xeon\textsuperscript{\textregistered} E5-2690 + one NVIDIA\textsuperscript{\textregistered} Tesla\textsuperscript{\textregistered} P100 per node)\cite{pizdaintweb}
and Summit (IBM\textsuperscript{\textregistered} AC922 with two 22-core Power9\textsuperscript{\texttrademark} + six NVIDIA\textsuperscript{\textregistered} Tesla\textsuperscript{\textregistered} V100 per node)\cite{summitweb}.

The purpose of this work is thus twofold: First, we collect data that we can actually use to improve Octo-Tiger by identifying potential bottlenecks.
To do so, we collect measurements running the production scenario for 40 time-steps both on Summit and Piz Daint, using 48 compute nodes on Piz Daint and 8 compute nodes on Summit (resulting in 48 GPUs in either case).
With those measurements, we can investigate the specific parts of Octo-Tiger on two distinct architectures in a distributed CPU/GPU run, providing insights into the different runtime behavior of Octo-Tiger regarding GPU-performance, CPU performance, and communication. For example, the communication seems to have a larger overhead on Piz Daint.

Second, we are showcasing the feasibility of APEX for large-scale runs,
collecting combined CPU and GPU performance measurements, showing that the
overhead introduced by the profiling itself is small enough to handle large
production-scale scenarios.
To this end, we are running the scenario both with and without APEX profiling
enabled for a scaling run on each machine, to see both the overhead on a few
compute nodes, and the runtime behavior when scaling to more nodes (with up to 2000 compute nodes on Piz Daint). 
Furthermore, we repeat these overhead measurements on Piz Daint for a CPU-only run to determine the performance impact of the NVIDIA\textsuperscript{\textregistered} CUDA\textsuperscript{\texttrademark} Profiling Tools Interface (CUPTI) which is used to collect the GPU performance data. 


To highlight the need for low profiling overhead, we can look at the short
runtimes for each time-step of Octo-Tiger: During the test runs on Summit, we
gather \num{5} GB of data during all eight runs. For each run \num{40} time-steps were executed. Each time-step takes
about \num{0.720325}\si{\second} on \num{128} Summit nodes and
consists of 6 iterations of the gravity solver, 3 iterations of the hydro
solver, and all required communication. On Piz Daint we collected \num{55} GB of data in total. Here, each time step on \num{2000} nodes took \num{0.7872625}\si{\second}. As time steps are serial in nature, these iterations are our smallest parallel
unit. As the time-steps only run for a few hundred milliseconds overall,
overheads introduced by the profiling can be very noticeable even if they only
take a few milliseconds in total as well.

The remainder of this work is structured as follows: In Section \ref{sec:related}, we take a
brief look at profiling solutions in other AMT frameworks. 
We then introduce the scientific scenario which we are simulating with
Octo-Tiger in Section \ref{sec:application}. This is the scenario we also used
to collect the profiling data by running it for 40 timesteps with APEX enabled.
Section \ref{sec:framework} in turn introduces Octo-Tiger itself, as well as the utilized software stack.
In Section \ref{sec:performance}, we show and discuss the collection of the
performance measurements for Octo-Tiger. In Section~\ref{sec:performance:improvements}, we test communication optimization and analyze the performance improvement. Finally, we conclude the paper in Section \ref{sec:conclusion}.

\section{Related work}
\label{sec:related}
For the related work, we focus on AMTs with distributed capabilities which are:
Legion~\cite{bauer2012legion}, Charm++~\cite{kale1993charm++},
Chapel~\cite{chamberlain2007parallel}, and UPC++~\cite{bachan2017upc++}. For a
more detailed review, we refer to~\cite{thoman2018taxonomy}.
Legion~\cite{bauer2012legion} provides \texttt{Legion Prof} for combined CPU and GPU
profiling which is compiled into all builds. Enabling the profiler produces log
files which can be viewed using the profiler.
Charm++~\cite{kale1993charm++} provides \texttt{Charm
debug}~\cite{debugger04} and the \texttt{Projections}
framework~\cite{TauICPP09} for performance analysis and visualization.
Chapel~\cite{chamberlain2007parallel} provides
\texttt{ChplBlamer}~\cite{zhang2018chplblamer} for profiling. UPC++ seems not
to have some dedicated tool for profiling, and any profiler supporting C++ is
recommended in their documentation.


Like HPX, nearly all of these runtimes provide a specialized tool that has been
designed to deal with the particular challenges of AMTs in general, and the
needs of the runtime system in particular.  APEX is a specialized tool in the case
of HPX, and provides similar measurement and analysis abilities of the
above tools, including flat profiling, tracing, sampling, taskgraphs/trees,
and concurrency graphs. In addition, APEX provides support for several
programming models/abstractions with or without HPX, including CUDA\textsuperscript{\texttrademark}, HIP,
OpenMP, OpenACC, Kokkos, POSIX threads, and C++ threads.
APEX does not provide analysis tools directly, but rather uses commonly
accepted formats and targets both HPC performance analysis tools
(ParaProf\cite{bell2003paraprof}, Vampir\cite{knupfer2008vampir},
Perfetto\cite{perfettoweb}) and standard data analysis tools (Python,
Graphviz\cite{graphvizweb}).

\section{Scientific application}
\label{sec:application}
Stellar mergers are mysterious phenomena that pack a broad range of physical processes into a small volume and a fleeting time duration. With the proliferation of deep wide-field, time-domain surveys, we have been catching on camera a vastly increased number of outbursts, many of which have been interpreted as stellar mergers. The best case, so far, of an observed merger is V1309 Sco, a contact binary identified using a recent survey database, the Optical Gravitational Lensing Experiment~\cite{Udalski1992}. Fortunately, not only the merger itself was observed, but archival data from other observing programs enabled the reconstruction of the light curve years before the merger. During the merger itself, the system brightness increased by 4 magnitudes, with a peak luminosity in the red visible light~\cite{Tylenda2011}. This complete record of observations has led to term V1309 Sco the “Rosetta Stone” of mergers. Previous attempts to model this merger included
semi-analytical calculations \cite{pejcha2014} and hydrodynamic simulations (e.g.,~\cite{nandez2014}).
However, the hydrodynamic simulations fail to adequately resolve the atmosphere, the rapid transition between the optically thick merger fluid and the optically thin, nearly empty space, surroundings of the simulated stellar material. To overcome this barrier, computational scientists intend to use the adaptive mesh-refinement hydrodynamics code Octo-Tiger. Using Octo-Tiger's dynamic mesh refinement, the simulations are able to resolve the atmosphere at a higher resolution than ever before. Simulation of the V1309 merger in high resolution provide greater insight into the nature of the mass flow and the consequential angular momentum losses.
In this paper, we analyze the performance of Octo-Tiger to identify potential bottlenecks in the combined CPU and GPU long-term production runs, where the atmosphere is maximally resolved. Analyzing the performance is essential at this stage since this model will serve as the necessary baseline for extending Octo-Tiger to include radiation transport to the V1309 model, as well as to other binary merger models.

\section{Software framework}
\label{sec:framework}

\subsection{C++ standard library for parallelism and concurrency}
HPX is the C++ standard library for parallelism and concurrency~\cite{Kaiser2020} and one of the distributed asynchronous many-task runtime systems) AMT. Other notable AMTs with distributed capabilities are: Uintah~\cite{germain2000uintah},
Chapel~\cite{chamberlain2007parallel}, Charm++~\cite{kale1993charm++}, Legion~\cite{bauer2012legion}, and PaRSEC~\cite{bosilca2013parsec}. However, according to~\cite{thoman2018taxonomy} HPX is one with a higher technology
readiness level. HPX's API is fully conforming with the recent C++ standard~\cite{cxx14_standard,cxx17_standard,cxx20_standard} which is the major difference to the other mentioned AMTs. For more details about HPX, we refer to~\cite{hpx_pgas_2014,Kaiser:2015:HPL:2832241.2832244,Heller2016,Kaiser2020}. In this work, HPX has two purposes: \textit{1)} the coordination of the synchronous execution of a multitude of heterogeneous tasks (both on CPUs and GPUs), thus managing local and distributed parallelism while observing all necessary data dependencies and \textit{2)}  as the parallelization infrastructure for executing CUDA\textsuperscript{\texttrademark}-kernels on the GPUs via the asynchronous HPX backend.

\subsection{APEX}

APEX~\cite{huck2015autonomic} is a performance measurement library for
distributed, asynchronous multitasking systems. It provides lightweight
measurements without perturbing high concurrency through synchronous and
asynchronous interfaces.
To support performance measurement in systems that employ user-level threading,
APEX uses a dependency chain in addition to the call stack to produce traces
and task dependency graphs.
The synchronous APEX instrumentation application programming interface (API)
can be used to add instrumentation to a runtime, library or application, and
includes support for timers and counters.
For measuring kernels on NVIDIA\textsuperscript{\textregistered} accelerated platforms, APEX is integrated with
the NVIDIA\textsuperscript{\textregistered} CUDA\textsuperscript{\texttrademark} Profiling Tools Interface~\cite{cuptiweb} that provides CUDA\textsuperscript{\texttrademark} host
callback and device activity measurements.
Similarly, on AMD accelerated platforms, APEX is integrated with
the Roctracer library~\cite{roctracerweb} providing HIP host
callback and device activity measurements.
In addition to timer measurements, the hardware and operating system are
monitored through an asynchronous measurement that involves the periodic or
on-demand interrogation of the operating system, hardware states, or runtime
states (e.g., CPU use, resident set size, memory ``high water mark'').
The NVIDIA\textsuperscript{\textregistered} Management Library interface~\cite{nvmlweb} provides periodic CUDA\textsuperscript{\texttrademark}
device monitoring to APEX, and the ROCm SMI API provides periodic HIP device
monitoring.
APEX has been extended to capture additional timers and counters related to
CUDA\textsuperscript{\texttrademark} device-to-device memory transfers, as well as tracking memory consumption
on both device and host when requested with the \lstinline{CUDAMalloc*}/\lstinline{cudaFree*} or
\lstinline{hipMalloc*}/\lstinline{hipFree*} API calls~\cite{wei2021memory}.

APEX supports tracing in both the OTF2 and Google Trace Events formats, but
for comparing results between platforms, profile data can be easier to work with.
To complement the profile data which collapses the time axis, APEX also captures task
and counter scatter plot data, indicating on the $x$ axis when the task started
or the counter was captured, and the $y$ axis contains the duration of the task
or the value of the counter. The tasks are sampled using a user-specified
fraction (default 1\%) whereas the counters are sampled at every value capture.
This data collection allows the application developer to capture a time
sequence of data without the file system overhead of a full event trace.

\subsection{Octo-Tiger}
\label{sec:framework-octotiger}
Octo-Tiger is a 3D adaptive mesh refinement (AMR) hydrodynamics finite volume code with Newtonian gravity designed specifically for the study of interacting stellar binaries~\cite{octotiger_apcs_2016}, \cite{marcello2021octo}. The AMR grid rotates  with the initial orbital frequency of the binary, which reduces inaccuracies due to numerical viscosity. The gravitational potential and force are computed using a modified version of the Fast Multipole Method that eliminates the gravitational field as a source of angular momentum conservation violation \cite{octotiger_fmm}. This enables Octo-Tiger to conserve energy and linear momentum to machine precision in the rotating frame. The astrophysical fluid is modeled using the inviscid Euler equations, which are solved with a finite volume central scheme~\cite{kurganov2000}.
The computational domain is based on a properly nested three-dimensional octree structure. Each node in the structure is an $N^3$ Cartesian sub-grid (usually, $N=8$), and may be further refined into eight child nodes, each containing its own $N^3$ sub-grid with twice the resolution of the parent. 

In~\cite{marcello2021octo}, an improved hydro solver for Octo-Tiger that includes a full three-dimensional reconstruction technique has been introduced.
The performance of this improved hydro solver (after porting it to GPUs) on Summit's ORNL has been tested as well as its accuracy~\cite{diehl2021octotigers}, showing a good GPU speedup (the exact amount depending on the chosen sub-grid size), and a greater accuracy in maintaining a rotating star in equilibrium than the old hydro solver. 
It has been fully benchmarked demonstrating superior angular momentum conservation and extreme scalability properties allowing it to compute larger problems in a shorter wall-clock time~\cite{marcello2021octo}. 
A convergence study in a real production run of a white dwarf merger has been also performed~\cite{diehl2021performance}.

Octo-Tiger's CUDA\textsuperscript{\texttrademark} implementation is worth elaborating on in a bit more detail here to understand the later performance results in Section~\ref{sec:performance}, in particular the short runtimes of the GPU kernels.
Usually, each of the compute kernels in the solver (both hydro solver and gravity solver) operates on a single sub-grid at a time with one CPU core. Multi-core usage is achieved by launching many of those methods on different sub-grids concurrently as HPX tasks.

In a GPU run, an issue arises where a single sub-grid does not provide enough work to utilize the entire GPU. However, the scheduler can simply launch multiple kernels on different sub-grids concurrently.
They are launched in different CUDA\textsuperscript{\texttrademark} streams, which are drawn from a pre-allocated pool to avoid the overhead of stream creation. Usually, we use $128$ streams per GPU.
As managing multiple CUDA\textsuperscript{\texttrademark} streams in each of the CPU threads might become unwieldy, the code uses an HPX-CUDA\textsuperscript{\texttrademark} integration. With it, HPX futures can be obtained from CUDA\textsuperscript{\texttrademark} kernel launches, allowing HPX to treat them as any other HPX task. Using this strategy, HPX can simply launch a CUDA\textsuperscript{\texttrademark} kernel, define subsequent post-processing tasks to be run once it is done, and then suspend the current HPX task that starts the kernel. The current CPU thread can then work on another task, potentially launching another GPU kernel in a different stream.

This implementation scheme has the implication that HPX runs many GPU kernels and CPU tasks concurrently, allowing the application to interleave GPU kernels, CPU tasks, CPU-GPU memory transfers, and inter-node communication seamlessly. Each of the tasks, both on CPU and GPU, has a rather short runtime as a result of only dealing with a single sub-grid at a time. For example, the individual GPU kernel execution time ranges from $100$ microseconds (less for smaller auxiliary kernels) to about $2$ milliseconds.

\section{Performance measurements}
\label{sec:performance}

Octo-Tiger's dependencies are listed in Table~\ref{tab:octo:dependencies} in the Artifact Description of the appendix. On both systems, we used Octo-Tiger's git hash (\textit{4c38f3bf}) as the baseline. However, we had to do some minor changes with regard to compilation, which do not affect the astrophysics kernels. Therefore, we used a slightly diverged git hash (\textit{b091fd26}) on Piz Daint and git hash (\textit{fd7faf5e}) on Summit. To quantify the overhead $o(n)$ introduced by APEX in percent, we define the metric
\begin{align}
    o(n) = \frac{comp\_apex(n)}{comp\_time\_no\_apex(n)} \cdot 100 - 100\,,
\label{eq:overhead}
\end{align}
where $n$ is the number of nodes, $comp\_time\_no\_apex(n)$ without APEX, and $comp\_apex(n)$ the computation time with APEX. Note that we only measure the time for the actual computation and not the IO.

\subsection{Distributed scaling}
\begin{figure}
    \centering
    \subfloat[\label{fig:distributed:scaling}]{
    \begin{tikzpicture}
    \begin{axis}[xlabel={\# nodes},ylabel={Cells processed per second},title={Distributed scaling (CPU + GPU)},grid,legend pos=north west,xmode=log,log basis x={2},xtick={1,2,4,8,16,32,64,128,256,512,1024,2048},ymode=log,log basis y={2},legend columns=2,legend style={at={(0.5,-0.2)},anchor=north}]
    \addplot[thick,mark=*,plot1] table [x expr=\thisrowno{0},y expr={512*(35855+35904)/2*40/\thisrowno{1}}, col sep=comma] {distributed-scaling-no-hyper.csv};
    \addplot[thick,mark=square*,plot1] table [x expr=\thisrowno{0},y expr={512*(35855+35904)/2*40/\thisrowno{1}}, col sep=comma] {distributed-scaling-no-apex-no-hyper.csv};
    \addplot[thick,mark=*,plot2] table [x expr=\thisrowno{0},y expr={512*(35855+35904)/2*40/\thisrowno{1}}, col sep=comma] {distributed-scaling-summit.csv};
    \addplot[thick,mark=square*,plot2] table [x expr=\thisrowno{0},y expr={512*(35855+35904)/2*40/\thisrowno{1}}, col sep=comma] {distributed-scaling-summit-no-apex.csv};
    \legend{Piz Daint with APEX,Piz Daint,Summit with APEX,Summit};
    \end{axis}
    \end{tikzpicture}
    }
    \\
    \subfloat[\label{fig:distributed:speedup}]{
    \begin{tikzpicture}
    \begin{axis}[xlabel={\# nodes},ylabel={Speedup},title={Speedup (CPU + GPU)},grid,legend pos=north west,xmode=log,log basis x={2},xtick={1,2,4,8,16,32,64,128,256,512,1024,2048},ymode=log,log basis y={2},legend columns=2,ymin=1]
    \addplot[thick,mark=*,plot1] table [x expr=\thisrowno{0},y expr={1915.98/\thisrowno{1}}, col sep=comma] {distributed-scaling-no-hyper.csv};
    \addplot[thick,mark=square*,plot1] table [x expr=\thisrowno{0},y expr={886.772/\thisrowno{1}}, col sep=comma] {distributed-scaling-no-apex-no-hyper.csv};
    \addplot[thick,mark=*,plot2] table [x expr=\thisrowno{0},y expr={1111.07/\thisrowno{1}}, col sep=comma] {distributed-scaling-summit.csv};
    \addplot[thick,mark=square*,plot2] table [x expr=\thisrowno{0},y expr={1034.04/\thisrowno{1}}, col sep=comma] {distributed-scaling-summit-no-apex.csv};
    \addplot[domain=1:2000]{x};
    \addplot [domain=1:2000,dashed,shift={(axis cs:4,1)},legend image post style={shift={(0,0)}}]{x};
    \legend{Piz Daint with APEX,Piz Daint, Summit with APEX,Summit,Optimal (Summit),Optimal};
    \end{axis}
    \end{tikzpicture}
    }
    \caption{Cells processed per second \protect\subref{fig:distributed:scaling} and speedup \protect\subref{fig:distributed:speedup}. On Piz Daint (\textcolor{plot1}{blue line}) we were able to use \num{4}, \num{8}, \num{16}, \num{32}, \num{64}, \num{128}, \num{256}, \num{512}, \num{1024}, \num{1400}, \num{1600}, \num{1800}, and \num{2000} nodes. On Summit (\textcolor{plot2}{violet line}) we used \num{1}, \num{2}, \num{4}, \num{8}, \num{16}, \num{32}, \num{64}, and \num{128} nodes. The speedup was obtained with respect to the smallest amount of nodes the scenario (\num{18} Million cells) fitted on. Note that for the runs with and without APEX a different time on the smallest nodes were used.}
\end{figure}
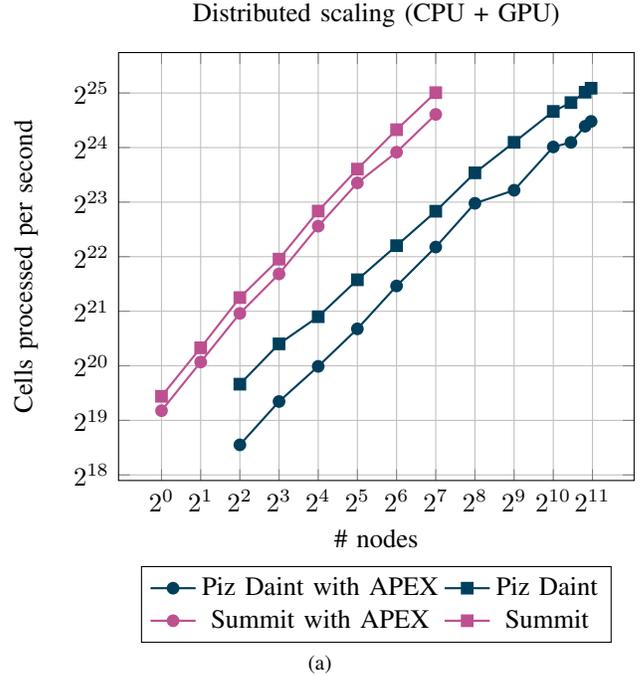
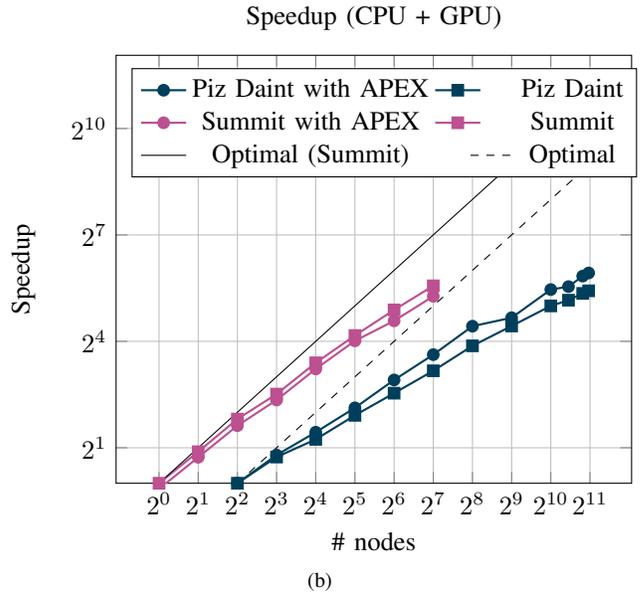

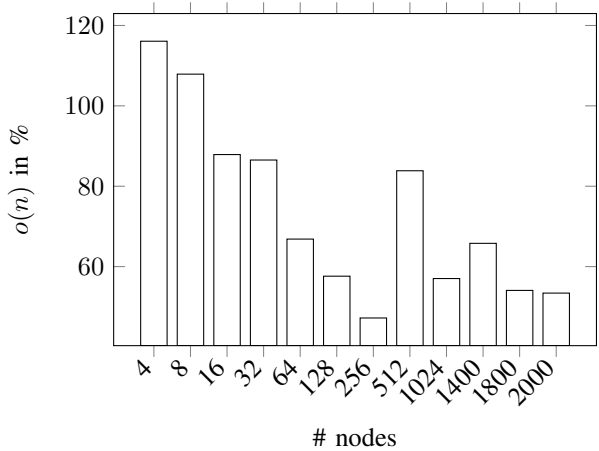
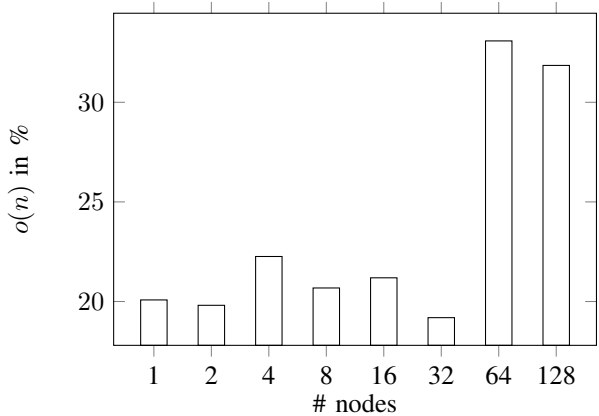
\begin{figure}[tb]
    \centering
        \subfloat[Piz Daint\label{fig:overhead:daint}]{
        \begin{tikzpicture}
\begin{axis} [ybar,height=6cm,width=8cm,xlabel=\# nodes,ylabel=$o(n)$ in \%,symbolic x coords={4,8,16,32,64,128,256,512,1024,1400,1800,2000}
,xtick=data,x tick label style={rotate=45,anchor=east},x label style={at={(axis description cs:0.5,-0.1)},anchor=north}]
\addplot [draw=black] coordinates {
    (4,116.08) 
    (8,107.89) 
    (16,87.88) 
    (32,86.52)
    (64,66.85)
    (128,57.62)
    (256,47.22)
    (512,83.85)
    (1024,57.03)
    (1400,65.79)
    (1800,54.08)
    (2000,53.41)
};
\end{axis}
\end{tikzpicture}
    }
    \\
    \subfloat[Summit\label{fig:overhead:summit}]{
    \begin{tikzpicture}
\begin{axis} [ybar,height=6cm,width=8cm,xlabel=\# nodes,ylabel=$o(n)$ in \%,symbolic x coords={1,2,4,8,16,32,64,128},xtick=data]
\addplot [draw=black] coordinates {
    (1,20.08) 
    (2,19.81) 
    (4,22.26) 
    (8,20.68)
    (16,21.19)
    (32,19.19)
    (64,33.08)
    (128,31.85)
};
\end{axis}
\end{tikzpicture}
    }
    \caption{Overhead $o(n)$ in Equation~\eqref{eq:overhead} for the runs on Piz Daint and Summit, respectively.}
    \label{fig:overhead}
\end{figure}

The largest V1309 scenario (\num{18} million cells) fitting on four CSCS's Piz Daint nodes was chosen for this scaling test. Note that the largest scenario for a single node was too small to scale out up to \num{2000} nodes with \num{2000} NVIDIA\textsuperscript{\textregistered} P\num{100} GPUs and \num{24000} CPU cores. However, due to the large amount of memory per node, the same scenario fits on one of ORNL's Summit node. On both machines, we executed the same runs with the APEX + CUDA\textsuperscript{\texttrademark} profiling on and off to investigate the overhead introduced by the profiling. Figure~\ref{fig:distributed:scaling} shows the cells processed per second for increasing amount of nodes. On Piz Daint the scaling was done using Octo-Tiger's new hydro solver~\cite{diehl2021octotigers} using a three-dimensional reconstruction scheme and scaling results using the old solver for a different scenario are shown here~\cite{daiss2019piz}. The \textcolor{plot1}{blue} lines show the scaling on Piz Daint. For both configurations, the scenario scales up to \num{2000} nodes. However, for \num{1400} Piz Daint nodes the problem size gets too small, and the scaling starts to flatten out. The overhead $o(n)$ in Equation~\ref{eq:overhead} for Piz Daint is shown in Figure~\ref{fig:overhead:daint}. The overhead is the largest on \num{4} nodes and seems to decline with increasing node counts.

The \textcolor{plot2}{violet} lines show the scaling on Summit. Here, again, the code scales well up to \num{128} nodes (using \num{768} NVIDIA\textsuperscript{\textregistered} V\num{100} GPUs overall) and the work is not sufficient for more nodes. The overhead $o(n)$ in Equation~\eqref{eq:overhead} on Summit is shown in Figure~\ref{fig:overhead:summit}. The overhead is less prominent on Summit.

We have seen an introduced overhead by the combined profiler on both systems. In a previous study using a different problem and the old version of the hydro module, the overhead introduced by pure APEX CPU profiling within HPX was around \num{1}\si{\percent}~\cite{diehl2021performance}. To verify if this still holds with the new hydro module and the V1309 scenario, we ran the V1309 scenario on Piz Daint with pure APEX CPU profiling. First, we observed scaling for the CPU kernels up to 2000 Piz Daint nodes. Second, we observed see that the difference is again around \num{1}\si{\percent} as in the previous study. It seems that the overhead is mostly introduced by enabling CUPTI measurement in APEX for both systems.  However, it seems that the overhead for smaller node counts on Piz Daint is larger, which needs to be investigated. Regardless, CUPTI provides varying levels of detail/support, and APEX should be refactored to enable the minimum amount of useful support by default, and allow the user to request additional details as needed.

\subsection{Profiling of Octo-Tiger}
\subsubsection{Setup}
Because APEX is directly integrated into the HPX runtime, annotating HPX actions (tasks) is simply a matter of providing an annotation for
the task when it is instantiated in the code.  Not all actions are annotated (anonymous lambdas launched through \lstinline{hpx::async()} calls, for example), but major operations in the Octo-Tiger code and in the HPX runtime are annotated.  When HPX is configured and built with APEX support, all annotations are provided to APEX, and APEX times the life cycle of each task.  HPX is designed to yield and resume tasks when resources are unavailable (futures, system calls, other dependencies), so APEX also provides the ability to yield (and resume) timing a task when it is not executing.  For GPU kernels, the kernel name is obtained using the instruction pointer address and debug information in the executable provided by the compiler.  

For the following Octo-Tiger profiling results, we use $48$ HPX localities (processes). 
As we use one HPX locality per GPU, this results in using $48$ compute nodes on Piz Daint and 8 compute nodes on Summit (as each node contains $6$ GPUs here).
This ensures that we use the same number of GPUs on both systems. We are still using the same V1309 scenario as before.

Below is a short description of the tasks and kernels of interest in Octo-Tiger.
The main CPU tasks include \lstinline{node_server::non-refined_step::compute_fluxes}: directs the computation of a single Runge-Kutta substep in the hydro and gravity solvers for a given sub-grid. It directs the GPU to compute hydrodynamic fluxes, corrects these fluxes on coarse-fine boundaries, works with other invocations of the same action to compute the global time-step size, computes hydrodynamic sources, then lastly updates the hydrodynamic variables and directs the GPU to update the gravitational variables;
\lstinline{local_step::execute_step}: directs the execution of an entire time-step for a given sub-grid. This involves multiple calls to \lstinline{node_server::nonrefined_step::compute_fluxes};
\lstinline{diagnostics_actions_type}: performs some measurements on the grid, for example, computing total mass on the grid, total angular momentum, and center of masses of each star (for binaries);
\lstinline{solve_gravity_action_type}: solves for gravity. This is called when an additional gravity solve is needed (other than what \lstinline{node_server::nonrefined_step::compute_fluxes} calls), such as after grid refinement;
\lstinline{check_for_refinement_action_type} flags each sub-grid in need of refinement;
\lstinline{regrid_gather_action_type} gathers information about the sub-grid structure and determines boundaries for the global decomposition;
\lstinline{regrid_scatter_action_type} uses the information computed by \lstinline{regrid_gather_action_type} to redistribute the sub-grids.

The GPU kernels include \lstinline{cuda multipole interactions kernel}: Computes the cell to cell interactions for the gravity solver in refined sub-grids (non-rho is without the angular momentum correction);
\lstinline{cuda p2p interactions kernels}: Computes the cell to cell interactions for the gravity solver in non-refined sub-grids;
Special case: \lstinline{cuda p2m interactions kernel}: Computes the cell to cell interactions for the gravity solver in non-refined sub-grids with refined neighbor sub-grids  (non-rho is without the angular momentum correction);
Special case: \lstinline{multipole root}: computes the remaining cell to cell interactions in the root sub-grid;
\lstinline{reconstruct cuda kernel} reconstructs the evolution variables using the PPM method, see \cite{diehl2021octotigers}; 
\lstinline{flux cuda kernel} the flux method computes the fluxes and the Newtonian quadrature to get the final flux.


\subsubsection{Results / Analysis}

In this section, we analyze the collected profiling results and draw conclusions regarding Octo-Tiger's performance on two different architectures. These insights are important for further optimization on the code. Some of them are straightforward and others are quite puzzling. Let us start with the insights on Summit:

\begin{figure}[tb]
    \centering
    \includegraphics[width=0.48\textwidth]{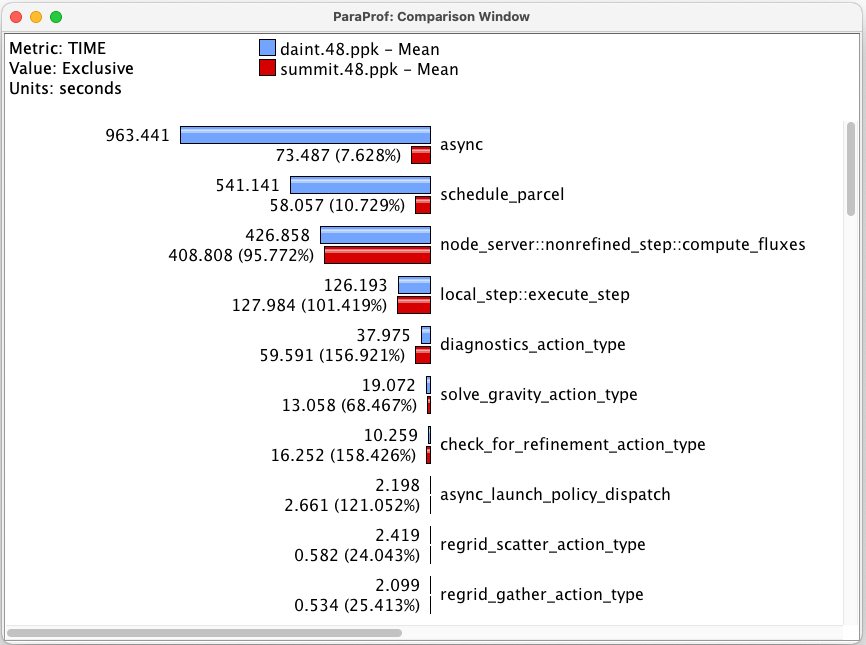}
    \caption{Comparison of top CPU tasks with TAU ParaProf.}
    \label{fig:comparison:cpu}
\end{figure}
\begin{figure}[tb]
    \centering
    \includegraphics[width=0.48\textwidth]{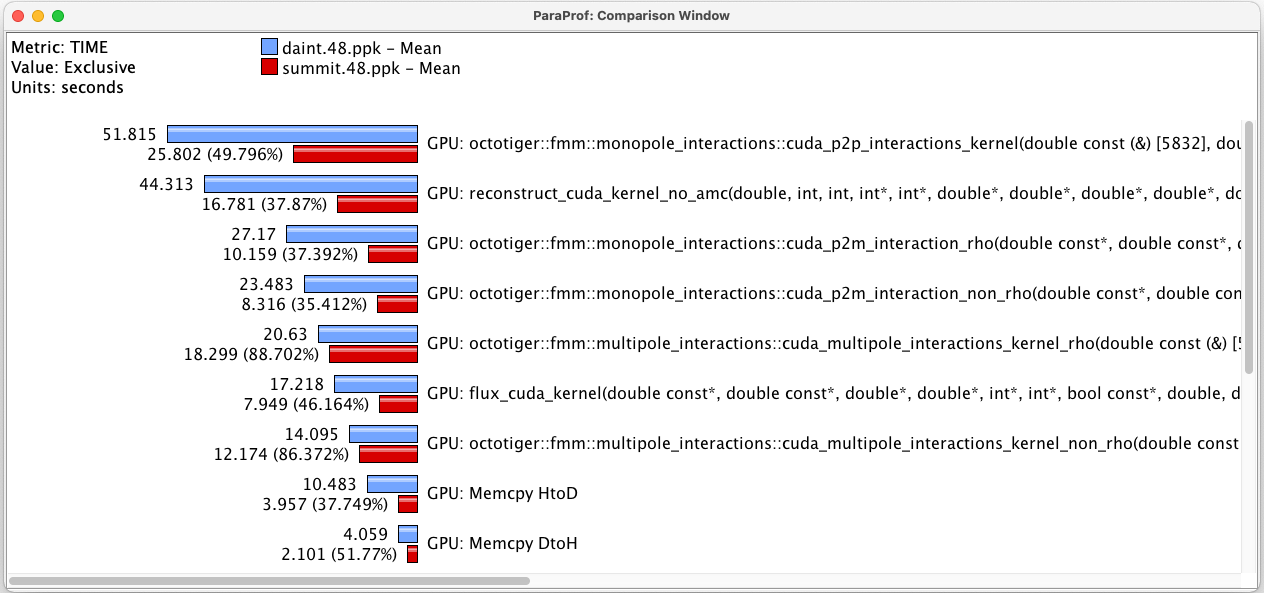}
    \caption{Comparison of top GPU kernels with TAU ParaProf.}
    \label{fig:comparison:gpu}
\end{figure}
\begin{figure*}
    \centering
    \includegraphics[width=\textwidth]{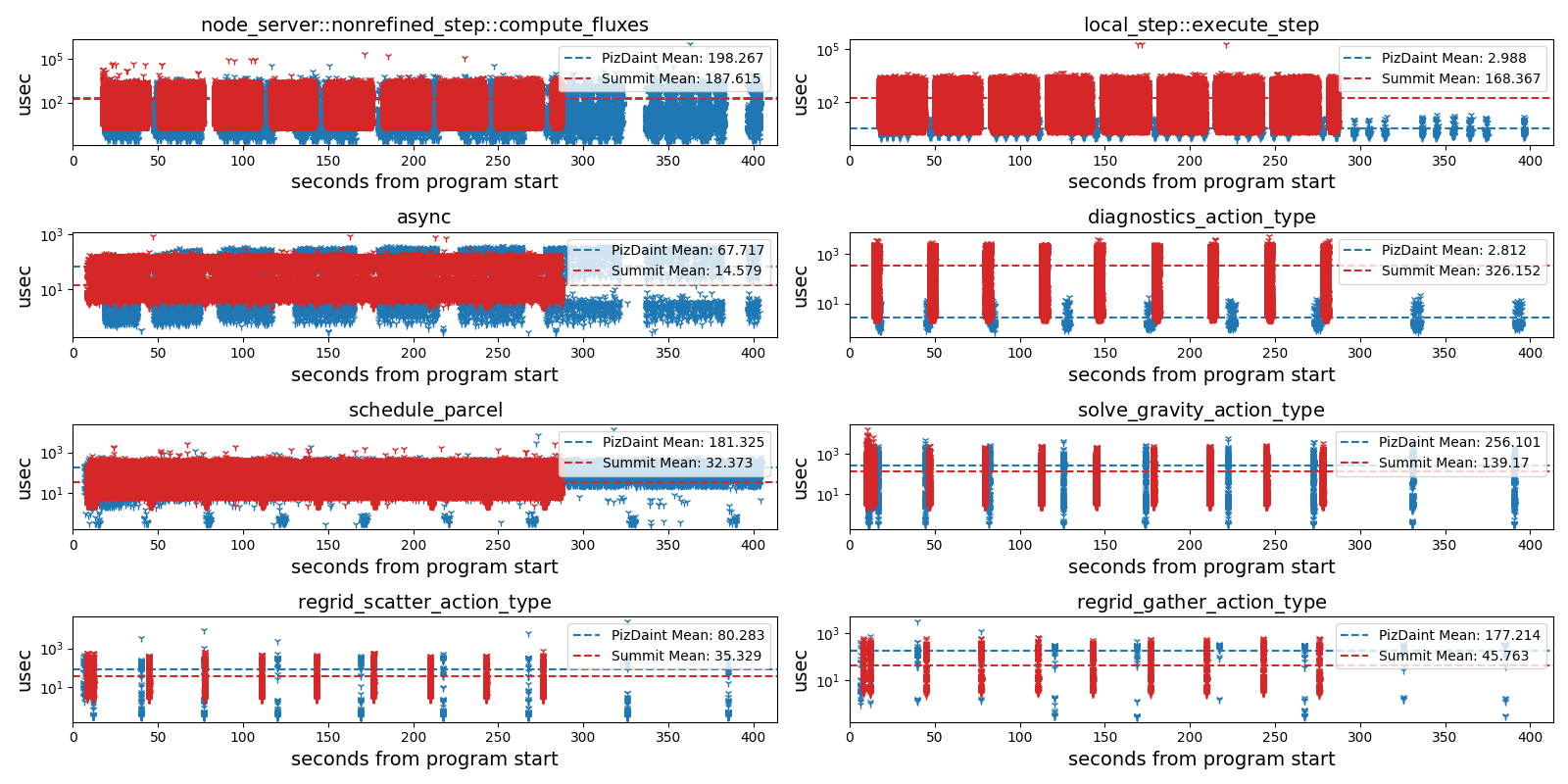}
    \caption{Comparison of sampled profiles of CPU tasks on 48 localities between Summit (\textcolor{red}{red}) and Piz Daint (\textcolor{blue}{blue}). The overall runtime on Summit is shorter, but interestingly, some CPU tasks took longer (per call) to execute on Summit than on Piz Daint.}
    \label{fig:cpu:scatterplot}
\end{figure*}
\begin{figure*}[tb]
    \centering
    \includegraphics[width=\textwidth]{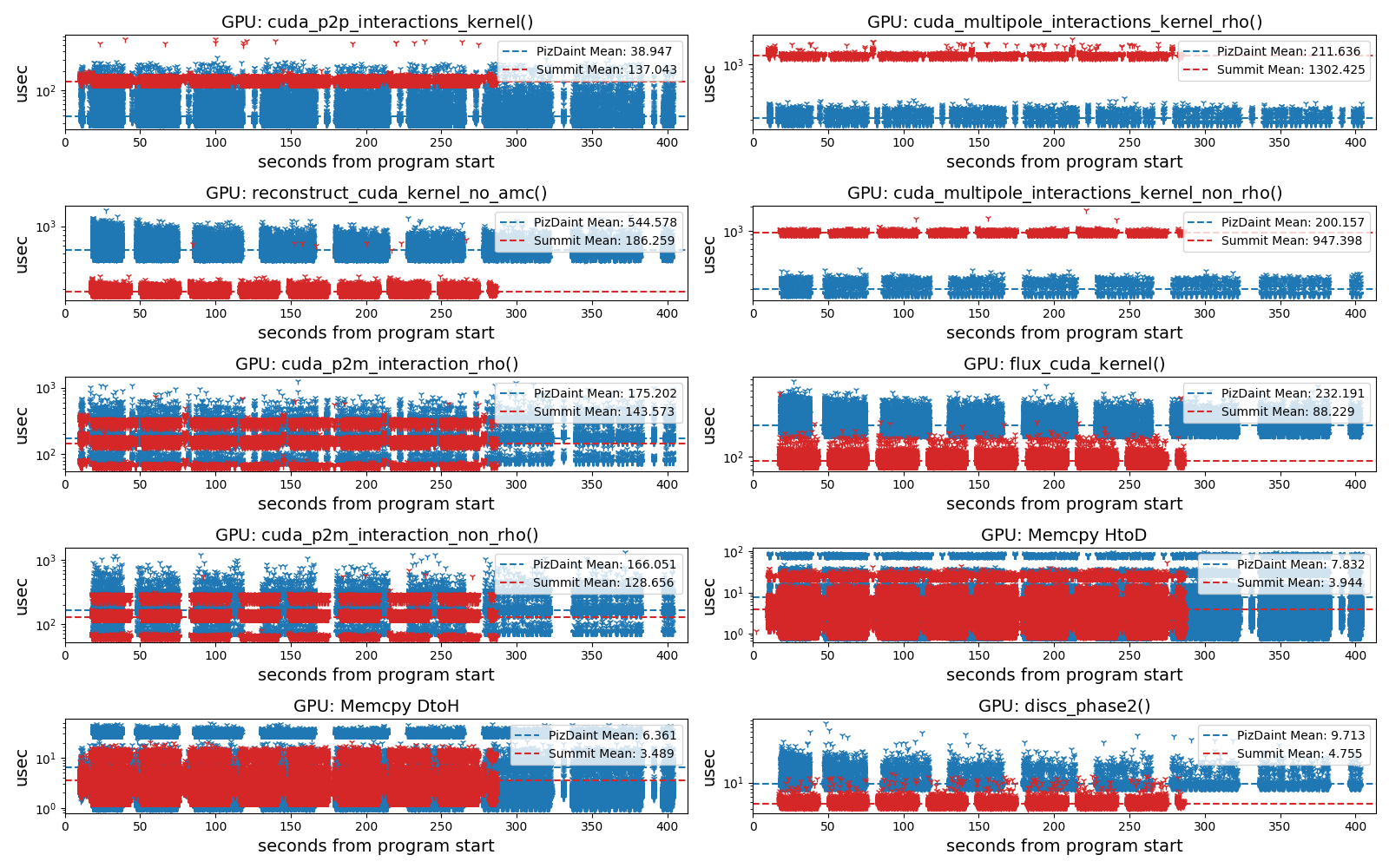}
    \caption{Comparison of sampled GPU kernels and memory transfers on 48 localities between Summit (\textcolor{red}{red}) and Piz Daint (\textcolor{blue}{blue}). Not surprisingly, in all cases, the performance of the V100s on Summit outperformed that of the P100s on Piz Daint. However, not all kernels saw a significant improvement in performance.}
    \label{fig:gpu:scatterplot}
\end{figure*}
\begin{enumerate}
    \item Figure~\ref{fig:comparison:cpu} shows that on Summit, the CPU tasks are not always faster with respect to Piz Daint. However, the overall computational time is lower due to the benefit of the newer GPUs. Looking at the scatterplot data for sampled timers in Figure~\ref{fig:cpu:scatterplot}, we see that while the average time \textit{per task} on Summit is slower for some tasks (\lstinline{compute_fluxes}, \lstinline{execute_step}), the aggregated mean profile is approximately the same. 
    \item Figure~\ref{fig:comparison:gpu} shows that on Summit, the GPU tasks are nearly always faster on  NVIDIA\textsuperscript{\textregistered} V100 GPUs. This is not surprising, since the GPU kernels are working better on the newer NVIDIA\textsuperscript{\textregistered} V100 than on the older NVIDIA\textsuperscript{\textregistered} P100. We found two exception, both the \lstinline{cuda_multipole_interactions_kernel_no_rho()} and \lstinline{cuda_interactions_kernel__rho()}  were faster on the NVIDIA\textsuperscript{\textregistered} P100. These tasks computes the monopole-multipole gravity interactions in the case that the leave nodes have different refinements. More investigation is needed to determine if it is possible to improve this compute kernel for the NVIDIA\textsuperscript{\textregistered} V100.
\end{enumerate}
On Piz Daint we gather the following insights:\\

    Unrelated to the application itself, we found some differences within the HPX run time system.  HPX provides the function \lstinline{hpx::async} to asynchronously launch functions and lambda functions. Figures~\ref{fig:comparison:cpu} and~\ref{fig:cpu:scatterplot} show that this operation's mean was $4.6\times$ more expensive on Piz Daint as on Summit. Here, the HPX main developers need to investigate this behavior.
    Figure~\ref{fig:comparison:cpu} also shows that an another more expensive HPX operation on Piz Daint was \lstinline{schedule_parcel} which mean was nearly $5.6\times$ higher as on Summit. A delay in the \lstinline{schedule_parcel} potentially happens when some \lstinline{hpx::future} is not ready, which \lstinline{schedule_parcel} depends on. 
Our primary focus here is distributed combined CPU and GPU profiling. We anticipate future research concentrating on the interpretation of these findings and major optimization. However, we will show one minor optimization.

\section{Performance improvements}
\label{sec:performance:improvements}
\begin{figure*}[tb]
    \centering
    \includegraphics[width=\textwidth]{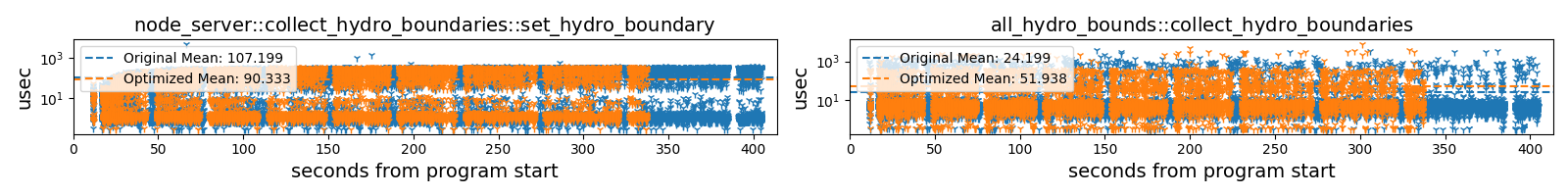}
    \caption{Comparison of sampled CPU and GPU kernels and 48 localities between the original code (\textcolor{blue}{blue}) and the optimized code (\textcolor{orange}{orange}) on Piz Daint.}
    \label{fig:scatterplot:optimized}
\end{figure*}
On Piz Daint the CPU tasks \lstinline{async} and \lstinline{schedule_parcle}, see Figure~\ref{fig:cpu:scatterplot}, took longer as on Summit.
We further noticed that this seems to be worse when running scenarios that included the Octo-Tiger hydro solver on Piz Daint when doing more tests.
Consequently, we took a look at the boundary communication within the Hydro module, and a way to reduce the overall number of messages. For sub-grids located on the same HPX locality, we now access the memory of these directly (foregoing HPX actions and temporary communication buffers) to fill the ghost-cells. This requires some more overhead within the \lstinline{node_server::collect_hydro_boundaries::set_hydro_boundary} method (and other associated methods) itself as we need to make sure that the results of those direct neighbors is up-to-date, which is done with local HPX promises/future pairs. This increases the mean time of said communication methods as they need to handle the promises, but reduces the overall HPX action calls and thus reduces the calls to schedule parcel. We gained a noticeable speedup on Daint (reduced total runtime from $400$ to $320$ seconds) and will be become helpful for other machines in the future as well.
This becomes evident in, Figure~\ref{fig:scatterplot:optimized} compares  the run on 48 Piz Daint nodes with the optimization enabled (\textcolor{orange}{orange}) and the previous run (\textcolor{blue}{blue}).
This does not address the root issue, which seem to be the longer runtimes for \lstinline{async} and \lstinline{schedule_parcle} on Daint, but it reduces the symptoms considerably and is furthermore a useful optimization for other machines as well.

Unfortunately, the test-bed allocation on Summit ended before we could finish implementing the optimization. Therefore, we could not show results here.

\section{Conclusion}
\label{sec:conclusion}
In this work, we have analyzed the overhead of performing combined CPU and GPU performance measurements with APEX in a large-scale HPX application distributed across up to \num{2000} compute nodes.

We have demonstrated that Octo-Tiger easily scales to that many nodes on Piz Daint with the APEX profiling enabled. Profiling is thus feasible for real-world production-size runs on high-performance systems equipped with GPUs.
However, we encountered a noticeable profiling overhead at scale (\num{52.408539388924}\% on 2000 Piz Daint nodes). This seems to be due to the GPU measurements with CUPTI, as a subsequent CPU-only run exhibited a smaller profiling overhead. Note that the overhead on Piz Daint with very few nodes is about two times higher and requires further investigation.
On Summit, there is a noticeable overhead at scale from profiling, too, but significantly less (\num{31.848570683336}\% on \num{128} Summit nodes). Overall, regarding the APEX profiling overhead, the distributed profiling with both CPU and GPU measurements works, scales up, and is ready to use; yet more investigation is needed to work on minimizing the overhead of the GPU measurements, if possible. 


While we focused on evaluating how suitable APEX is for these large-scale, distributed analyses, there are some interesting results regarding Octo-Tiger itself as well.
It is notable that the speedup of the average GPU kernel runtime from a P100 to a newer V100 varies a lot between the kernels. There are many factors that may influence the average kernel runtime difference between the devices: For instance, going to the V100, there is an increase in the number of Streaming Multiprocessors (SM), an increase in L1 cache available per SM, an increase in global memory bandwidth and a slight increase in clock speed. This does not even take into account the fact that we compile for different architectures.

Considering that we use concurrent kernel execution via $128$ CUDA\textsuperscript{\texttrademark} streams per GPU to achieve device utilization (as outlined in Section~\ref{sec:framework-octotiger}), it is unlikely that the increased numbers of SMs in the V100 has a major impact on the average kernel execution time (as those would rather facilitate more concurrent kernel execution). Thus, it seems more likely that the speedup is due to a combination of the other factors, the exact speedup depending on what is currently limiting the kernel. The larger speedups indicate that the kernels benefit from the larger L1 cache available, however, determining the exact cause is subject of future work, especially as the kernels are currently still undergoing changes. However, the results here give us an idea which kernels need more attention during this process, particularly when targeting older architectures, thus helping us to steer our development focus.

Beyond the GPU results, the profiling uncovered that there are some crucial methods that run significantly slower on Piz Daint than on Summit, e.g. \lstinline{schedule_parcel}. We consequently optimized the communication of the hydro solver to alleviate this issue. Using the APEX measurements we could identify the parts of the code which benefited from the optimizations. This shows the need for distributed performance measurements on a production system.
Of course, while we focused on the usability of APEX for these kinds of analyses in this work and fixed some of the issues, the uncovered issues still need to be further investigated and addressed. Consequently, we will examine these remaining issue in future work, hopefully further improving the runtime of future simulations with Octo-Tiger.


A radiation module for Octo-Tiger is currently being implemented by its developers, and it is in the testing phase. Our performance analysis of the current modules will be 
crucial to estimate the performance impact of the new module prior to its inclusion. 
Including radiation in the simulations of V1309 together with resolving the star atmosphere at a higher resolution than ever before will enable one to self-consistently compute the light curve and directly compare it with the observed one of V1309.
If one is able to accurately reproduce the light curve of the "Rosetta Stone of mergers", it will be possible to reliably simulate the outburst light curves of other mergers.
%

\section*{Acknowledgment}
\label{sec:acknowledgement}
{\footnotesize This work was supported by a grant from the Swiss National
Supercomputing Centre (CSCS) under project ID s1078. This research used resources of the Oak Ridge Leadership Computing
Facility, which is a DOE Office of Science User Facility supported under Contract DE-AC05-00OR22725.
The APEX work was supported by the Scientific Discovery through Advanced Computing (SciDAC) program funded by U.S. Department of Energy, Office of Science, Advanced Scientific Computing Research (ASCR) under contract DE-SC0021299.}

\section{Supplementary materials}
The build scripts are available on GitHub\footnote{\url{https://github.com/STEllAR-GROUP/OctoTigerBuildChain}} and the input files are available on Zenodo~\cite{marcello_2021_5213015}.

\cleardoublepage
\newpage

\bibliographystyle{./bibliography/IEEEtran}
\bibliography{./bibliography/IEEEabrv,./bibliography/IEEEexample}

\end{document}